\documentclass[structabstract]{raa}
\usepackage{graphicx,times}
\usepackage{natbib,amssymb}
\usepackage{epsfig}
\usepackage{float}

\providecommand{\bm}[1]{\mbox{\boldmath$#1$\unboldmath}}

\begin{document}
   \title{Is the X-ray pulsating companion of HD 49798 a possible type Ia supernova progenitor?}

   \volnopage{ {\bf 2015} Vol.\ {\bf X} No. {\bf XX}, 000--000}
   \setcounter{page}{1}

   \author{D.-D. Liu \inst{1,2,3}
          \and
          W.-H. Zhou \inst{3,4,5}
          \and
          C.-Y. Wu \inst{1,2,3}
          \and
          B. Wang \inst{1,2}
          }

   \institute{Yunnan Observatories, Chinese Academy of Sciences, Kunming 650216, China
          {\it liudongdong@ynao.ac.cn; wangbo@ynao.ac.cn}\\
          \and
          Key Laboratory for the Structure and Evolution of Celestial Objects, Chinese Academy of Sciences, Kunming 650216, China
          \and
          University of Chinese Academy of Sciences, Beijing 100049, China\\
          \and
          National Astronomical Observatories, Chinese Academy of Sciences, Beijing 100012, China\\
          \and
          Yunnan Minzu University, Kunming 650031, China\\
              }

   \date{Received ; accepted}

\abstract {HD 49798 (a hydrogen depleted subdwarf O6 star) with its massive white dwarf (WD) companion has been suggested to be a progenitor candidate of type Ia supernovae (SNe Ia). However, it is still uncertain whether the companion of HD 49798 is a carbon-oxygen (CO) WD or an oxygen-neon (ONe) WD. A CO WD will explode as an SN Ia when its mass grows approach to Chandrasekhar mass, while the outcome of an accreting ONe WD is likely to be a neutron star. We followed a series of Monte Carlo binary population synthesis approach to simulate the formation of ONe WD + He star systems. We found that there is almost no orbital period as large as HD 49798 with its WD companion in these ONe WD + He star systems based on our simulations, which means that the companion of HD 49798 might not be an ONe WD. We suggest that the companion of HD 49798 is most likely a CO WD, which can be expected to increase its mass to the Chandrasekhar mass limit by accreting He-rich material from HD 49798. Thus, HD 49798 with its companion may produce an SN Ia in its future evolution.
 }

\titlerunning{Is the X-ray pulsating companion of HD 49798 a possible type Ia supernova progenitor?}

\authorrunning{D.-D. Liu et al.}

\maketitle


\section{Introduction} \label{1. Introduction}

Type Ia supernova (SN Ia) explosions are among the most luminous phenomena in the Universe, and play an important role in astrophysics. Due to the remarkable uniformity of their high luminosities, SNe Ia are successfully used as standard cosmological distance indicators. It has been suggested that the Universe is expanding at an increasing rate through the observation of SNe Ia, which reveals the existence of dark energy (e.g., Riess et al. 1998; Perlmutter et al. 1999). Furthermore, SN Ia explosions are also relatively important for galactic chemical evolution for the reason that a great amount of iron can be produced during this process (e.g., Greggio \& Renzini 1983; Matteucci \& Greggio 1986), and they are also accelerators of cosmic rays (e.g., Fang \& Zhang 2012). However, SN Ia progenitors and their explosion mechanisms are still uncertain, which may have an influence on the accuracy of being the role of distance indicators (e.g., Podsiadlowski et al. 2008; Howell et al. 2011; Liu et al. 2012; Wang \& Han 2012; Wang et al. 2013; Hillebrandt et al. 2013; Maoz et al. 2014).

A theoretical consensus has been reached that SNe Ia originate from thermonuclear runaway explosions of carbon-oxygen white dwarfs (CO WDs) in binaries (see Hoyle \& Fowler 1960; Nomoto et al. 1997). When a CO WD increases its mass approach to the Chandrasekar mass limit in a close binary, an SN Ia would be produced. However, the key issue is the uncertainty of the companion star. There are two competing progenitor models which are widely accepted, i.e., the single-degenerate model and the double-degenerate model. In the single-degenerate model, a CO WD accretes material from a non-degenerate companion to get its mass increased. The companion in this model could be a main sequence (MS) star, a subgiant star, a red giant (RG) star or a heluim (He) star (e.g., Whelan \&Iben 1973; Nomoto et al. 1984; Li \& van den Heuvel 1997; Langer et al. 2000; Han \& Podsiadlowski 2004; Han \& Podsiadlowski 2006; Meng et al. 2009; Xu \& Li 2009; Wang et al. 2009a, 2010; Ablimit et al. 2014; Chen H.-L. et al. 2014; Geier et al. 2015). In the double-degenerate model, the merger of two CO WDs produces an SN Ia; the merger is due to the loss of orbital angular momentum driven by gravitational wave radiation (e.g., Tutukov \& Yungelson 1981; Iben \& Tutukov 1984; Webbink 1984; Han 1998; Chen et al. 2012).

The CO WD + He star channel is an emerging variant of the single-degenerate model which can naturally account for the missing of hydrogen lines in the most spectra of observed SNe Ia. Wang et al. (2009a) systematically studied the CO WD + He star channel and derived the space parameters for producing SNe Ia. They found that SNe Ia from this channel may have a contribution to SNe Ia with short delay times. In the observations, a number of massive WD + He star binaries have been found, e.g., V445 Pup, HD 49798 with its companion, CD$-$30$^{\circ}$\,11223 and KPD 193$+$2752, etc. All of these binaries are possible candidates of the SN Ia progenitors (e.g., Geier et al. 2007, 2013; Kato et al. 2008; Woudt et al. 2009; Wang \& Han 2010; Mereghetti et al. 2011). Among these observed WD + He star systems, HD 49798 with its massive compact companion is one of the relatively well studied binaries composed of a bright subdwarf star whose hydrogen has been depleted (Jaschek \& Jaschek 1963) and a fast X-ray pulsar source which is a massive WD with a short spin period (Israel et al. 1997).

HD 49798/RX J0648.0-4418 is a single spectroscopic binary which has been investigated extensively.
Jaschek \& Jaschek (1963) first classified HD 49798 as a subdwarf O6 star and obtained the variations of its radial velocity.
The period of this binary was suggested to be 1.548\,day (see Thackeray 1970; Stickland \& Lloyd 1994).
A soft X-ray emission source was first found by Einstein Observatory in the position of HD 49798 (see Simon et al. 1979).
According to the detailed data from Roentgen Satellite, Israel et al. (1997) found that the X-ray source in the HD 49798/RX J0648.0-4418 is very soft and has a high-energy excess, and that the X-ray source has a pulsating period of 13.2s.
Bisscheroux et al. (1997) argued that the companion is a WD but not a neutron star based on its low X-ray luminosity (see also Hamann et al. 1981; Mereghetti et al. 2011).
Bisscheroux et al. (1997) suggested that the X-ray emission of this binary is from wind accretion process of the compact object, in which the accretion rate from the stellar wind of HD 49798 is $\sim 10^{-10}-10^{-11} \rm \,M_{\odot}/yr$.
They also speculated that HD 49798 is at He-shell burning stage, which might be the explanation for its high luminosity.
Recently, Mereghetti et al. (2009) observed the object with the Newton X-ray Multimirror Mission satellite during the X-ray eclipse.
They gave the parameters of this binary: $R_{\rm He}=1.45\pm0.25 \,R_{\odot}$, $M_{\rm He}=1.50\pm0.05 \,M_{\odot}$ for HD 49798 and $M_{\rm WD}=1.28\pm0.05 \,M_{\odot}$ for the WD companion.

Wang \& Han (2010) recently performed a detailed binary evolutionary calculation for HD 49798/RX J0648.0-4418. The temperature and luminosity of HD 49798 derived from their calculations are consistent with that of observations (see Fig.\,1 of Wang \& Han 2010). Their work indicates that HD 49798 will fill its Roche lobe after $\sim 4\times \rm 10^4 \rm \, yr$, and the WD companion of HD 49798 will increase its mass to Chandrasekar mass by accreting He-rich material. However, it is still unknown which kind of WD the companion of HD 49798 is. If the compact companion is a CO WD, it will grow to be an SN Ia when the carbon is ignited in the center (or off-center) of the WD (e.g., Nomoto 1982). If the companion of HD 49798 is an ONe WD, Ne and Mg in the WD start to capture electrons when the WD grows to Chandrasekhar mass. During the deflagration of O and Ne, too little energy could be released to give rise to an explosion of the whole WD (Miyaji et al. 1980). Driven by the subsequent electron capture, the outcome of the WD should be a neutron star with fast spin but not an SN Ia (e.g., Canal et al. 1980; Nomoto \& Kondo 1991).

The purpose of this article is to examine whether the companion of HD 49798 is an ONe WD or not via a binary population synthesis (BPS) method. The BPS numerical code and the input physics are described in Section 2. In Section 3, we present the results of our calculations. The discussion and conclusions are provided in Section 4.

\section{Methods}

By using Hurley's rapid binary evolution code (Hurley et al. 2000, 2002), we performed a series of Monte Carlo simulations to study the formation of ONe WD + He star systems. In this work, an ONe WD is formed by the envelope loss of a  thermal pulsing asymptotic giant branch (TPAGB) star with $M_{\rm up} \le M \le M_{\rm ec}$, while a CO WD is formed by the envelope loss of a TPAGB star with $M < M_{\rm up}$ (see Hurley et al. 2000), where $M$, $M_{\rm up}$ and $M_{\rm ec}$ are the initial mass of the AGB star, the minimum mass undergoing non-degenerate C ignition and the minimum mass of an AGB star avoiding electron capture on Ne and Mg in its core, respectively (see table 1 of Pols et al. 1998). In each simulation, $1\times10^7$ primordial binary systems are included. The current parameters of HD 49798 and its WD companion has been provided by Mereghetti et al. (2009, 2011), that is, the masses of the He star and the WD are $1.50\pm0.05 \,M_{\odot}$ and $1.28\pm0.05 \,M_{\odot}$, respectively. The orbital period of this binary is 1.548 day (Thackeray 1970). Moreover, HD 49798 is a slightly evolved He star which has not filled its Roche lobe yet; the He star will fill its Roche lobe after about $4\times \rm 10^4 \rm \, yr$ (see Wang \& Han 2010), which means that the masses and the orbit period of the systems are similar to the parameters at the beginning of the formation of the WD + He star system. Thus, if the companion of HD 49798 is an ONe WD, the parameters of this binary are more likely to be located within the contours of ONe WD + He star systems determined by Monte Carlo BPS simulations.

\begin{figure}
   \begin{center}
\includegraphics[width=14.8cm,angle=0]{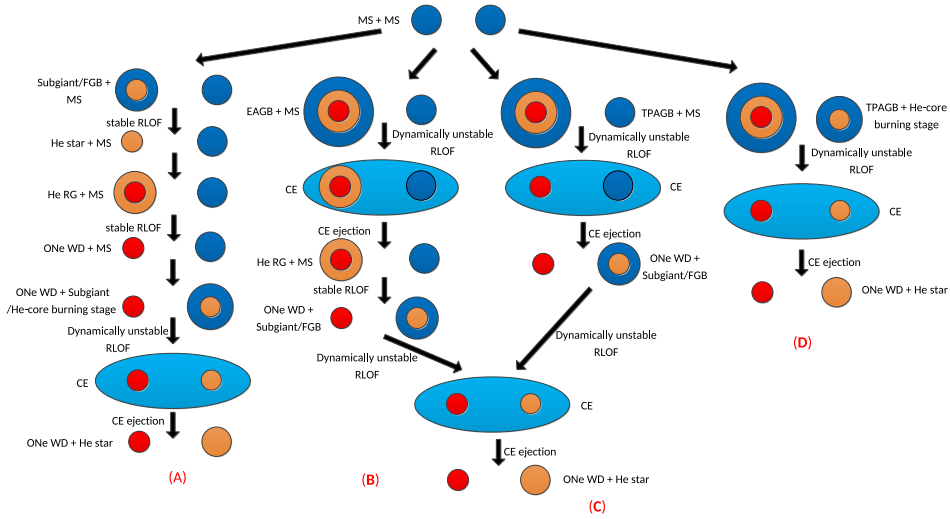}
 \caption{Four binary evolutionary scenarios to produce ONe WD + He star systems.}
   \end{center}
\end{figure}

According to the evolutionary phase when the primordial primary first fills its Roche lobe, there are four scenarios which can produce ONe WD + He star systems (see Fig.\,1), as follows:

Scenario A: When the primordial primary is a subgiant or first giant branch (FGB) star, it fills its Roche lobe. At this stage, the Roche lobe overflow (RLOF) is stable. At the end of the RLOF, the primordial primary becomes a He star. Subsequently, the He star continue to evolve and will fill its Roche lobe again at the He RG stage. As a result, an ONe WD + MS system is produced after RLOF. Then the MS star continues to evolve and will fill its Roche lobe at the subgiant or He-core burning phase. A common envelope (CE) may be formed due to the dynamically unstable mass transfer. If the CE ejection happens, an ONe WD + He star system would be produced.

Scenario B: When the primordial primary first fills its Roche lobe at the early asymptotic giant branch (EAGB) stage, the mass transfer is dynamically unstable, leading to the formation of a CE. After the CE ejection, the primary becomes a He RG star and the orbital separation decays sharply. Then the primary fills its Roche lobe again. In this case, the RLOF is stable. After the He-shell is exhausted, an ONe WD + MS system is produced. The MS star continues to evolve and may fill its Roche lobe at the Subgiant or FGB stage. A CE will be formed for the reason of the dynamically unstable mass transfer. If the CE can be ejected, an ONe WD + He star will be produced.

Scenario C: The primordial primary first fills its Roche lobe at the TPAGB stage and a CE may be formed due to the dynamically unstable mass transfer. After the CE ejection, an ONe WD + Subgiant/FGB star system is formed. Subsequently, the binary evolution is similar to that of the scenario B above and an ONe WD + He star system will be formed.

Scenario D: The primordial primary fills its Roche lobe when it is a TPAGB star and the primordial secondary is at the He-core burning phase. In this case, the mass transfer is dynamically unstable, leading to the formation of a CE. If the CE ejection happens, the primary turns to be an ONe WD and the secondary becomes a He star, i.e., an ONe WD + He star system is produced.

Some basic assumptions in our BPS simulations are listed as follows: (1) The initial mass function described by Miller \& Scalo (1979) is adopted. The primordial primary mass ranges from $0.1 \,M_{\odot}$ to $100 \,M_{\odot}$. (2) The primordial secondary mass is determined by the initial primordial primary mass and initial mass-ratio. For simplicity, a uniform mass-ratio distribution is adopted (e.g., Mazeh et al. 1992; Goldberg \& Mazeh 1994). (3) The distribution of the separations of the initial orbit is constant in log($a$) ($a$ is orbital separation), and all stars are set as members of binary systems with a circular orbit. While a boundary for close and wide binaries is set to be $10 \,R_{\odot}$ (Han et al. 1995). (4) The abundance of solar metallicity $Z=0.02$ is adopted. (5) The process of RLOF is calculated by the method described in Tout et al. (1997). (6) Stable mass transfer process presented by Webbink (1985) is adopted. (7) The output of the CE stage is calculated by the standard energy equation (e.g., Webbink 1984), in which there are two uncertain parameters, i.e., the CE ejection efficiency ($\alpha_{\rm CE}$) and a stellar structure parameter ($\lambda$). Following the work of Wang et al. (2009b), we simply combined these two parameters as a single free parameter (i.e., $\lambda\alpha_{\rm CE}$). In this work, we set $\lambda\alpha_{\rm CE}$ to be 0.5, 1.0, 1.5 and 2.0 to examine its influence on the production of ONe WD + He star systems.\footnote{Recent studies showed that the stellar structure parameter, $\lambda$, may be bigger during some particular stages (e.g., Xu \& Li 2010; Zuo \& Li 2014). Thus, we considered an extreme case with $\lambda\alpha_{\rm CE}=2.0$.}

\section{Results}

\begin{table}
\begin{center}
 \caption{Initial parameters of binaries which can evolve to ONe WD + He star systems. The simulation contains $1\times10^7$ primordial sample binaries, in which we set $\lambda\alpha_{\rm CE}=1.0$. Notes: $M_{\rm 1,0}=$ initial mass of the primordial primary; $M_{\rm 2,0}=$ initial mass of the primordial secondary; $P_{\rm 0}=$ initial orbital period of the primordial binary; Number = the number of ONe WD + He star systems produced from each scenario.}
 \label{tab1}
   \begin{tabular}{cccccc}
\hline \hline
  Scenario & $M_{\rm 1,0}$ & $M_{\rm 2,0}$ & $P_{\rm 0}$ & Number\\
 & $(M_\odot)$ & $(M_\odot)$ & (day)\\
\hline
  (A) & $8.0-11$ & $2.5-10$ & $2.0-960$ & $2469$\\
  (B) & $6.0-9.0$ & $2.5-8.5$ & $420-1600$ & $1704$\\
  (C) & $6.0-8.5$ & $2.5-7.5$ & $1400-6750$ & $2027$\\
  (D) & $6.0-8.5$ & $6.0-8.5$ & $1500-6000$ & $227$\\
\hline \label{1}
\end{tabular}
\end{center}
\end{table}

We conducted a series of Monte Carlo BPS approach to simulate the formation of ONe WD + He star systems. In Table\,1, we presented the initial parameters of binaries for those four scenarios in Fig.\,1 and gave the number of ONe WD + He star systems produced from each scenario. From this table, we can see that these four scenarios are mainly distinguished by the difference of initial orbital periods, while the distinguishing factor between scenario C and scenario D is the initial mass ratio.

\begin{figure}
   \begin{center}
\includegraphics[width=10.8cm,angle=0]{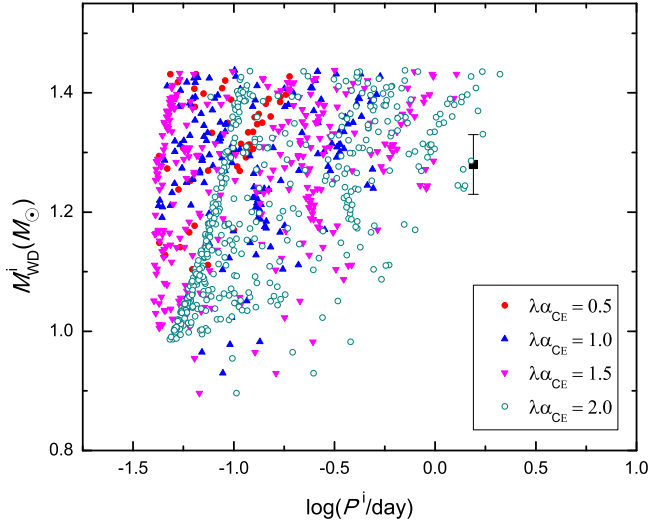}
 \caption{The initial orbital period and WD mass ($\rm log \, P^i$ and $M^{\rm i}_{\rm WD}$) distribution of ONe WD + He star systems based on BPS simulations, in which He star mass is constrained between 1.45 and 1.55 $\,M_{\odot}$. The dots, triangles, inverted triangles and open circles represent the cases with $\lambda\alpha_{\rm CE}=0.5$, 1.0, 1.5 and 2.0, respectively. The black square with error bar represents HD 49798/RX J0648.0-4418.}
   \end{center}
\end{figure}

Based on the four cases of simulations with different CE ejection parameters (i.e., $\lambda\alpha_{\rm CE}=0.5$, 1.0, 1.5 and 2.0), we totally obtained about 28000 ONe WD + He star binaries. From these binary systems, we picked out binaries with the He star mass between 1.45 and $1.55\,M_{\odot}$ which is consistent with the possible mass range of HD 49798. Fig.\,2 shows the distribution of 937 ONe WD + He star systems in the initial orbital period and WD mass ($\rm log \, P^i$-$M^{\rm i}_{\rm WD}$) plane with the constraint of the He star mass. From this figure, we can see that as the value of $\lambda\alpha_{\rm CE}$ increases, the orbital separation tends to be wider and more ONe WD + He systems would be produced. This is because, for a larger value of $\lambda\alpha_{\rm CE}$ the ejection of CE will release less orbital energy and be easier to happen. Note that, the upper boundaries is determined by the existence of Chandrasekar mass limit for ONe WDs, while the lower boundaries are constrained by the minimum mass of WDs for the ignition of carbon to form ONe WDs. The ONe WD + He star systems beyond the left boundaries have their He star masses lower than the mass range of HD 49798, whereas larger than that mass range beyond the right boundary. Considering the cases with $\lambda\alpha_{\rm CE}=0.5$, 1.0 and 1.5, we can see that no ONe WD + He star system has a separation (orbital period) as large as HD 49798/RX J0648.0-4418 for the same masses of He stars. While for the extreme case with $\lambda\alpha_{\rm CE}=2.0$, HD 49798/RX J0648.0-4418 is located on the boundary of the contour of the obtained ONe WD + He star systems. However, the probability is very low. We note that the extreme case with $\lambda\alpha_{\rm CE}=2.0$ may not be physical in our simulations. Thus, we speculated that HD 49798 with its companion might not be an ONe WD + He star system.

In Fig.\,3, we present a more comprehensive distribution of orbital periods of ONe WD + He star systems without the constraints of He star mass. From this figure, we can see that almost all the ONe WD + He star systems distribute between 0.032 and 1.0 day. As to the orbital period of HD 49798/RX J0648.0-4418 (1.548 day), for the cases with $\lambda\alpha_{\rm CE}=0.5$, 1.0 and 1.5, it is still difficult to produce such a wide ONe WD + He star system even without the constraints of the He star mass; the orbital period of HD 49798/RX J0648.0-4418 can be reproduced by the extreme case with the value of $\lambda\alpha_{\rm CE}=2.0$, but the probability is very low. Note that there are two peaks in the curves with $\lambda\alpha_{\rm CE}=1.5$ and 2.0, which are produced from different forming scenarios (see Fig.\,1); binaries near the left peak are formed from the scenario A, whereas the right peak are from all those four scenarios.

\begin{figure}
   \begin{center}
\includegraphics[width=10.8cm,angle=0]{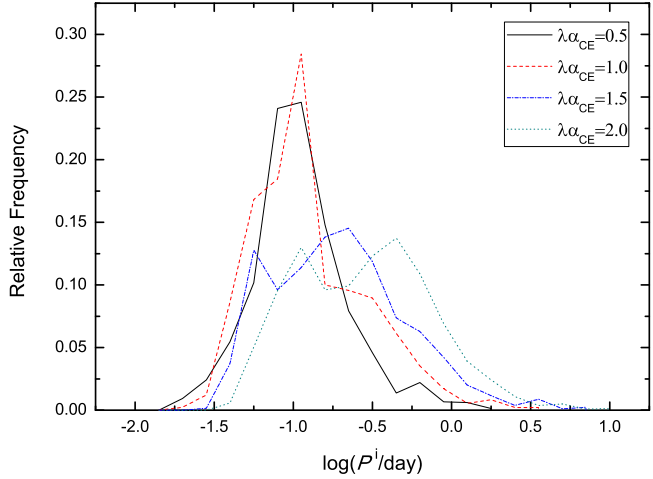}
 \caption{Distribution of the orbital periods of ONe WD + He star systems with different values of $\lambda\alpha_{\rm CE}$, in which the mass of He stars are not restricted. The solid, dashed, dash-dotted and dotted lines represent the cases with $\lambda\alpha_{\rm CE}=0.5$, 1.0, 1.5 and 2.0, respectively. Here, every case is normalized to be 1.}
   \end{center}
\end{figure}

\section{Discussion and conclusions} \label{6. DISCUSSION}

In this article, we found that there is almost no orbital period as large as HD 49798/RX J0648.0-4418 in these ONe WD + He star systems, which means that the companion of HD 49798 may not be an ONe WD under our assumptions.\footnote{The orbital period of HD 49798/RX J0648.0-4418 can be reproduced by the CO WD + He star model, which has the range of $0.01-5.62\, \rm day$.} Thus, the ultimate fate of the binary can be constrained to a certain extent.
By assuming the compact companion is a CO WD, Wang \& Han (2010) recently made an evolutionary investigation of this binary. In their calculations, the optically thick wind model is adopted and the mass accumulation efficiency of He-shell flash is from Kato \& Hachisu (2004). Wang \& Han (2010) suggested that about $6\times \rm 10^4 \, \rm yr$ later, the WD companion of HD 49798 will increase its mass to $1.4 \,M_{\odot}$ and an SNe Ia may be produced, leaving a surviving He star as massive as $1.1817 \,M_{\odot}$. Furthermore, Mereghetti et al. (2009) suggested that the companion of HD 49798 may be a rapidly rotating WD, which would lead to a CO WD more massive than Chandrasekhar mass (e.g., Yoon \& Langer 2005; Chen \& Li 2009). From Fig.\,2 of Wang \& Han (2010) we can see that, the WD companion of HD 49798 could eventually increase its mass to $1.62 \,M_{\odot}$ when the differential rotation cannot be maintained. Thus, we speculate that the companion of HD 49798 may evolve to an overluminous SNe Ia.

However, if the companion of HD 49798 is a CO WD, a question would be raised that how such a massive CO WD like the companion of HD 49798 could be formed with almost no mass transfer from the He star before it fills its Roche lobe. HD 49798/RX J0648.0-4418 is considered to be the production of a CE ejection and spiral-in process (e.g., Israel et al. 1997; Bisscheroux et al. 1997). Iben \& Tutukov (1993) claimed that HD 49798 was produced from an $8-9 \,M_{\odot}$ progenitor which had filled its Roche lobe before the He-core burning stage. They thought that HD 49798 is currently a He-core burning star (see also Kudritzki \& Simon 1978). In contrast, Bisscheroux et al. (1997) argued that the progenitor of HD 49798 is an EAGB star before the CE is formed and currently HD 49798 is at He-shell burning stage with a CO core in the center. Furthermore, Wang \& Han (2010) simulated the evolution of the binary HD 49798/RX J0648.0-4418 and found that the mass of the CO core is about $0.79 \,M_{\odot}$ at the current position of HD 49798.

The compact companion of HD 49798 might also be a hybrid CONe WD. When the convective boundary mixing is taken into account, a super AGB star may evolve to form a hybrid CONe core with a relatively large unburned CO core (about $0.2 \,M_{\odot}$), which surrounded by an ONe zone (greater than $0.85 \,M_{\odot}$) and a thin CO layer on the surface (Denissenkov et al. 2013). By considering a series of convective Urca shell flashes and some different mixing assumptions, Denissenkov et al. (2015) found that an explosive carbon ignition would be reached when hybrid CONe WDs approach to the Chandrasekhar mass, leading to a low peak luminosity SNe Ia because of the low carbon to oxygen radio. Wang et al. (2014) gave the birthrate of SNe Ia of CONe WD + He star scenario using a BPS approach. Furthermore, Chen et al. (2014) suggested that hybrid CONe WDs could be as massive as $1.3 \,M_{\odot}$ in an extreme case. Thus, the companion of HD 49798, which is as massive as $1.28 \,M_{\odot}$, may also be a hybrid CONe WD. However, it is still unclear that what is the relationship between the mass of CO core and ONe regions and what is the smallest CO core mass for thermonuclear explosion happens. Moreover, the carbon-burning rate, which is still an uncertain parameter, plays an important role in determining the biggest mass of hybrid CONe WDs (see Chen M.-C. et al. 2014). The uncertainties of hybrid CONe WDs make it doubtful that whether such a massive WD like the companion of HD 49798 could be a hybrid CONe WD.

We also note that, it is still being argued about the predicted ability of the BPS method. Toonen et al. (2014) recently compared four different BPS codes and found that the differences of their results are not because of numerical effects, but due to different Monte Carlo simulation assumptions (e.g., initial mass function, initial mass ratio distribution, orbital eccentricity distribution, etc). Wang et al. (2009b) compared some initial parameters for producing SNe Ia through the CO WD + He star model. Their work indicates that the initial assumptions adopted in this article may result in a maximum SN Ia birthrate. In this work, we performed four sets of simulations. The results presented here show that a larger value of $\lambda\alpha_{\rm CE}$ could produce wider and more ONe WD + He star systems. We also presented an extreme case with $\lambda\alpha_{\rm CE}=2.0$, and found that HD 49798 with its WD companion may be reproduced in the ONe WD + He star model with an extreme value of $\lambda\alpha_{\rm CE}$, but the probability is very low.

Aside from HD 49798 with its X-ray pulsating companion, another well studied WD + He star system is V445 Pup which is the first detected He nova (Ashok \& Banerjee 2003; Kato \& Hachisu 2003). The mass of the He star in V445 Pup is about $1.2-1.3 \,M_{\odot}$ derived from the pre-outburst luminosity of the binary (Woudt et al. 2009). Kato et al. (2008) fitted the light curve of the binary and estimated that the mass of the WD companion is larger than $1.35 \,M_{\odot}$. They also found that almost half of the material acceted from the He star remains on the surface of the WD. Recently, Goranskij et al.(2010) declared that the orbital period of V445 Pup might be 0.65\,day. Morever, the WD companion is more likely a CO WD but not a ONe WD since no Neon line was detected in the ejected nebula (e.g., Woudt \& Steeghs 2005). Thus, V445 Pup is also a possible candidate of SN Ia progenitors.

In this article, we performed a series of Monte Carlo BPS simulations with different values of CE ejection parameters. A number of ONe WD + He star systems were obtained and the companion of HD 49798 is really difficult to be an ONe WD under our simulations. If we exclude the case of an ONe WD, the companion of HD 49798 may be a CO WD or a CONe hybrid WD, both of which can be expected to evolve to an SN Ia. Because of the rapidly rotation of the companion, it is possible to produce an over-luminos SN Ia. If the companion of HD 49798 is a CO WD, it would be a challenge for the current binary evolution theory to explain how such a massive CO WD was formed. We hope that this work can stimulate more observations on the WD + He star systems so that more detailed studies of WD + He star channel can be proceeded.

\begin{acknowledgements}
We acknowledge useful comments and suggestions from the referee.
This study is supported by
the National Basic Research Program of China (973 program, 2014CB845700),
the National Natural Science Foundation of China (Nos 11322327 and 11390374),
and the Natural Science Foundation of Yunnan Province (Nos 2013FB083 and 2013HB097).

\end{acknowledgements}

\label{lastpage}

\end{document}